\documentclass[aps,twocolumn,floatfix]{revtex4} 

\usepackage{amsmath,amsfonts,amssymb,graphicx,color,times,bbm}

\newcommand{\id}{{\mathbbm{1}}}
\renewcommand{\vec}[1]{\boldsymbol{#1}}
\newcommand{\me}{\mathrm{e}}
\newcommand{\mi}{\mathrm{i}}
\newcommand{\md}{\mathrm{d}}

\definecolor{green}{rgb}{0.2,.7,0.4}

\begin{document}

\title{Interaction dependent temperature effects in Bose-Fermi mixtures in optical lattices}

\author{M.\ Cramer} 

\affiliation{Institut f\"ur Theoretische Physik, Albert-Einstein
Allee 11, Universit\"at Ulm, Germany}
   
\begin{abstract}
We present a quantitative finite temperature analysis of a recent experiment with Bose-Fermi mixtures in optical lattices, in which 
the dependence of the coherence of bosons on the inter-species interaction was analyzed. Our
theory reproduces the characteristics of this dependence and suggests that intrinsic temperature effects play an important
role in these systems. Namely, under the assumption that the ramping up of the optical lattice is an isentropic process, adiabatic temperature changes of the mixture
occur that depend on the interaction between bosons and fermions. Matching the entropy of two regimes---no lattice on the one hand and deep lattices on the other---allows us to compute the temperature in
the lattice and the visibility of the quasi-momentum distribution of the bosonic atoms, which we compare to the experiment.
We briefly comment on the remaining discrepancy between theory and experiment, speculating that it may in part be attributed to the dependence of the Bose-Fermi scattering length on the confinement of the atoms.
\end{abstract}

\maketitle

\date{\today}

\section{Introduction}
Ultracold atoms in optical lattices are, due to the available impressive control over system parameters, 
ideal candidates for ``quantum simulators" that mimic condensed matter systems \cite {cond_mat_sim1, cond_mat_sim2}. 
We have already seen them display the transition from a superfluid to a Mott insulator \cite{MI-sf},
Fermi surfaces have been observed \cite{FermiSurface}, and, recently, the finite temperature phase diagram for bosonic superfluids in an optical lattice has been obtained experimentally \cite{quantum_simulator}.
Multi-component mixtures, among them mixtures of bosonic and fermionic atoms, offer a variety of additional quantum phases of matter. Charge-density waves \cite{cdw}, supersolids \cite{supersolids1,supersolids2}, exotic superfluid \cite{exotic_sf1,exotic_sf2} and Mott-insulator phases \cite{lewenstein_bfh1,our_bfh1,mering_MI} have been predicted
 and we will certainly see experimental signatures of these in the near future.
Of course, temperature plays a prime role for such quantum simulators and its influence needs to be understood or, better yet, be under control. 
But, even just determining the temperature in an optical lattice is an extremely difficult task \cite{quantum_simulator}. Thermometry methods for such systems without
lattices are however well established. Hence, under the assumption that the lattice is ramped up adiabatically (which is usually a good approximation and was also recently confirmed for bosons in optical lattices \cite{quantum_simulator}), i.e., without changing the entropy, one may
infer about the temperature in the lattice using entropy-matching methods. Not only does this hold the opportunity for thermometry in the lattice but also offers the possibility to further cool the atoms \cite{adiabatic_bosons}.

In this work, we study interaction-dependent temperature effects occurring in  Bose-Fermi mixtures under this assumption
of raising the optical lattice being isentropic. We compare our results to the visibility of the quasi-momentum distribution, which was obtained in the experiment in Ref.~\cite{experiment} for a $^{87}$Rb-$^{40}$K mixture. By matching the entropy of two very different regimes (with and without lattice), we are able to take all experimental parameters (such as the anisotropic trapping potential, the number of particles and the lattice parameters) into account, leaving no free parameters in our theory. Our results show that these effects have a significant influence on the coherence
of the bosonic atoms and depend quite strongly on the interaction between the two species, in agreement with the experiment. Hence, we are faced with a situation in which intrinsic adiabatic temperature effects play a dominant role and, as we argue below, have already been observed experimentally.
\begin{figure}
\hspace{-0.3cm}\includegraphics[width=0.85\columnwidth]{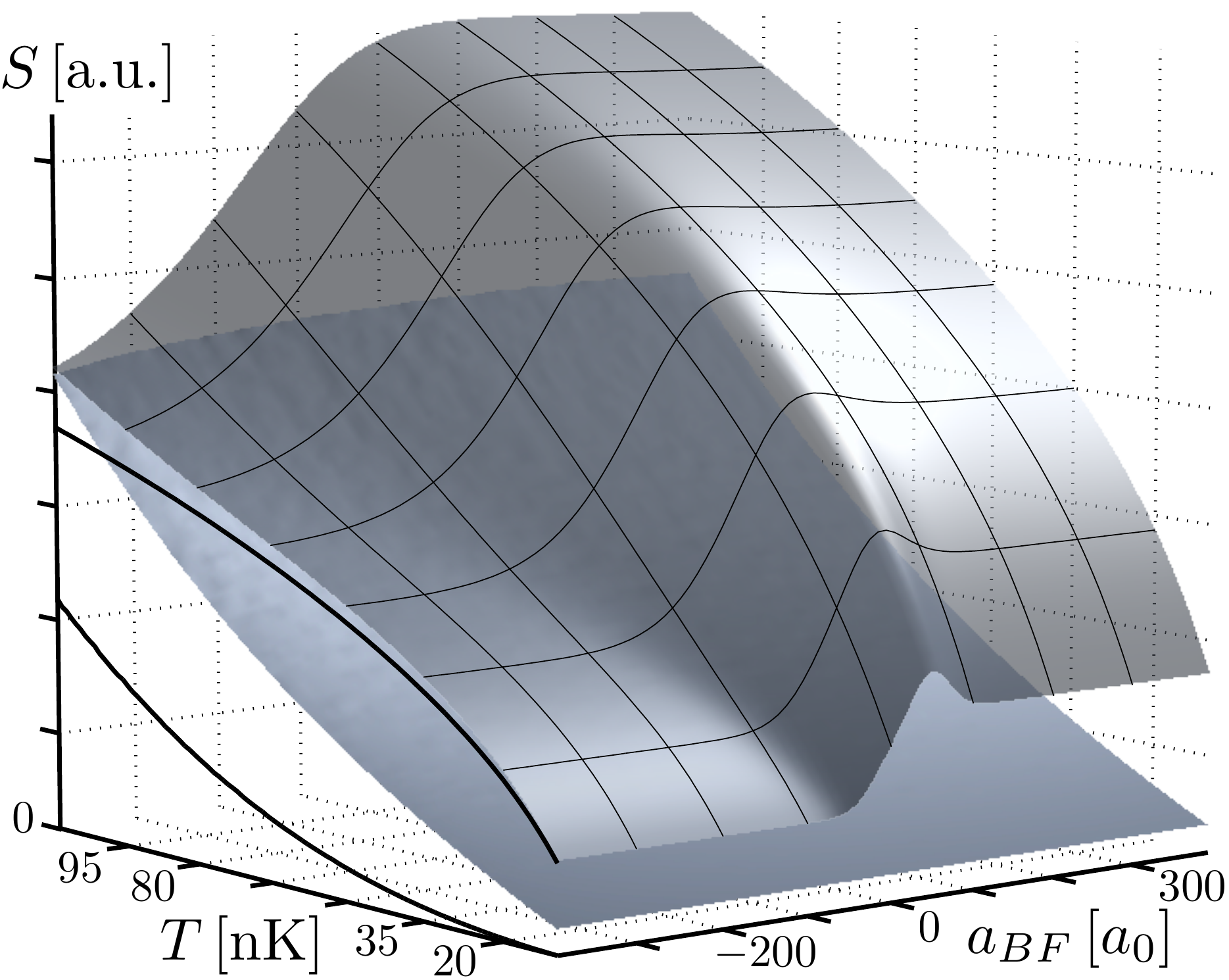}
\caption{\label{entropy}
Entropy as a function of temperature and Bose-Fermi scattering length $a_{BF}$ ($\propto g_{BF}\propto V$, the Bose-Fermi interaction) in units of the Bohr radius $a_0$ with (translucent surface) and without (opaque surface) a twelve recoil energies deep optical lattice. Bold lines at $a_{BF}=-400$ show the same for a purely bosonic system, for which the entropies are of course independent of $a_{BF}$ (upper line: including the lattice, lower line: without lattice). The mixture consists of $4\times 10^5$ $^{87}$Rb and $3\times 10^5$ $^{40}$K atoms and trap parameters are as in the experiment in Ref.~\cite{experiment}, see footnote \cite{parameters} for details.}
\end{figure}

More specifically, we calculate the entropy as a function of the temperature in the absence of the lattice invoking the Hartree-Fock-Bogoliubov-Popov mean-field approximation for the bosons, including the fermions in a self-consistent mean field interaction. To describe the mixture in the lattice,
we use the single-band Bose-Fermi-Hubbard Hamiltonian and calculate the entropy as a function of the temperature  for a deep lattice perturbatively. For both regimes, we assume that the mixture is in thermal equilibrium such that we can assign one temperature to it.
This results in temperature-entropy diagrams as in Fig.~\ref{entropy} and allows us to obtain the temperature in the lattice $T_f$ as a function of the initial temperature $T_i$ by matching the corresponding entropies, see Fig.~\ref{Tfin}. The resulting adiabatic heating or cooling of these isentropic processes was already analyzed for purely fermionic \cite{adiabatic_fermion_non_interacting} and bosonic \cite{adiabatic_bosons_proper,adiabatic_bosons} systems,
a Fermi gas of atoms that can pair into molecules via a Feshbach resonance \cite{adiabatic_molecules}, and Bose-Fermi mixtures \cite{adiabatic1} as in this work. For fixed inter-particle interaction, loss of bosonic coherence due to presence of fermions
was observed in the experiments in Refs.~\cite{bfh_exp1,bfh_exp2} and  attributed to intrinsic temperature effects in Refs.~\cite{bfh_exp1,kollath,adiabatic1} (see, however, also Refs.~\cite{self_trapping,qmc_bfh_1D}, in which different explanations were put forward). The adiabatic assumption was recently confirmed experimentally for bosons \cite{quantum_simulator}. These studies show that $T_f$ can depend strongly on the system parameters, prime among them the inhomogeneity introduced by the trapping potential, and hence
realistic descriptions should take all of them into account.
To the best of our knowledge, the present analysis is the
first to be directly comparable to the experiment over the full range of inter-species interactions and to fully take the experimental situation into account.

As our scheme to obtain $T_f$ is similar to the one used in Ref.~\cite{adiabatic1} (note however, that here we do not take geometrical averages of the trapping frequencies but take the full anisotropic situation into account and do not invoke the local density approximation in the lattice), we merely  outline it in the two subsequent Sections~\ref{no_lattice} and \ref{lattice}. Being equipped with $T_f$, we calculate the visibility of the quasi-momentum distribution within thermal perturbation theory and compare it to the experiment \cite{experiment} in Section~\ref{section_visi}, resulting in Fig.~\ref{visi}.

\section{No lattice}
\label{no_lattice}
We start from the microscopic model of a mixture of bosonic atoms of mass $m_B$ and fermionic atoms of mass $m_F$ subject to respective trapping potentials $V_{B}(\vec{r})$, $V_{F}(\vec{r})$, and interacting via contact interactions parametrized by the s-wave  scattering lengths $a_{BB}$ (Bose-Bose interaction)
and $a_{BF}$ (Bose-Fermi interaction),
\begin{equation}
\label{microscopic}
\begin{split}
\hat{H}&=\!\!\sum_{S=B,F}\int\!\!\md\vec{r}\,\hat{\Psi}^\dagger_S(\vec{r})\big[\tfrac{-\hbar^2\vec{\nabla}^2}{2m_S}+V_S(\vec{r})-\mu_S\big]\hat{\Psi}_S(\vec{r})\\
&\hspace{1cm}+\tfrac{g_{BB}}{2}\!\!\int\!\!\md\vec{r}\,\hat{\Psi}^\dagger_B(\vec{r})\hat{\Psi}^\dagger_B(\vec{r})\hat{\Psi}_B(\vec{r})\hat{\Psi}_B(\vec{r})\\
&\hspace{1cm}+g_{BF}\!\!\int\!\!\md\vec{r}\,\hat{\Psi}^\dagger_B(\vec{r})\hat{\Psi}_B(\vec{r})\hat{\Psi}^\dagger_F(\vec{r})\hat{\Psi}_F(\vec{r}).
\end{split}
\end{equation}
Invoking the Hartree-Fock-Bogoliubov-Popov mean-field approximation \cite{stringari_journal,stringari_review} for the bosons and including the fermions in a self-consistent mean-field approximation \cite{fermi_mean_field1,fermi_mean_field2}, yields the following set of coupled equations, which we solve self-consistently with an iterative numerical scheme (see, e.g., Ref.~\cite{adiabatic1} for more details). (i) The finite temperature Gross-Pitaevskii equation in the Thomas-Fermi approximation for the condensed atoms,
\begin{equation}
n_0(\vec{r})=\max\big\{0,\tfrac{\mu_B-V_B(\vec{r})-g_{BF}m(\vec{r})}{g_{BB}}-2n_T(\vec{r})\big\},
\end{equation}
where $m(\vec{r})$ is the fermionic density, $n_T(\vec{r})$ the density of thermal bosons, 
$g_{BB}=4\pi\hbar^2a_{BB}/m_B$, $g_{BF}=2\pi\hbar^2a_{BF}(m_B+m_F)/(m_Bm_F)$, and
the chemical potential $\mu_B$ controls the total number of bosons $N_B=\int\!\md\vec{r}\,n(\vec{r})$, $n(\vec{r})=n_0(\vec{r})+n_T(\vec{r})$. (ii) The thermal contribution
\begin{equation}
n_T(\vec{r})=\int\!\!\tfrac{\md\vec{p}}{(2\pi)^3}\Big[\tfrac{u_+(\vec{p},\vec{r})+u_-(\vec{p},\vec{r})}{\me^{\beta\varepsilon(\vec{p},\vec{r})}-1}+u_-^2(\vec{p},\vec{r})\Big],
\end{equation}
where $\beta=1/(k_BT)$ denotes the inverse temperature and the Bogoliubov amplitudes and quasi-particle spectrum are given by
\begin{equation}
\begin{split}
u_{\pm}(\vec{p},\vec{r})&=\tfrac{\frac{\hbar^2\vec{p}^2}{2m_B}+V_B(\vec{r})-\mu_B+2g_{BB}n(\vec{r})+g_{FB}m(\vec{r})}{2\varepsilon(\vec{p},\vec{r})}\pm\tfrac{1}{2},\\
\varepsilon(\vec{p},\vec{r})&=\big[\tfrac{\hbar^2\vec{p}^2}{2m_B}+V_B(\vec{r})-\mu_B+2g_{BB}n(\vec{r})\\
&\hspace{2.3cm}+g_{FB}m(\vec{r})\big]^2-g_{BB}^2n_0^2(\vec{r}).
\end{split}
\end{equation}
(iii) The fermionic density in local density approximation
\begin{equation}
m(\vec{r})=\int\!\!\tfrac{\md\vec{p}}{(2\pi)^3}\tfrac{1}{\me^{\beta\delta(\vec{p},\vec{r})}+1},
\end{equation}
where
\begin{equation}
\delta(\vec{p},\vec{r})=\tfrac{\hbar^2\vec{p}^2}{2m_F}+V_F(\vec{r})-\mu_F+g_{BF}n(\vec{r})
\end{equation}
and $\mu_F$ controls the total number of fermions $N_F=\int\!\md\vec{r}m(\vec{r})$.

After convergence, we are in a position to compute the entropy of the mixture in thermal equilibrium, $S(T)/k_B=\int\md\vec{p}\md\vec{r}[s_B(\vec{p},\vec{r})+s_F(\vec{p},\vec{r})]/(2\pi)^3$, with individual contributions given by
\begin{equation}
\begin{split}
s_B(\vec{p},\vec{r})&=\tfrac{\beta\varepsilon(\vec{p},\vec{r})}{\me^{\beta\varepsilon(\vec{p},\vec{r})}-1}-\log\big(1-\me^{-\beta\varepsilon(\vec{p},\vec{r})}\big),\\
s_F(\vec{p},\vec{r})&=\tfrac{\beta\delta(\vec{p},\vec{r})}{\me^{\beta\delta(\vec{p},\vec{r})}+1}+\log\big(1+\me^{-\beta\delta(\vec{p},\vec{r})}\big).
\end{split}
\end{equation}
For the parameters \cite{parameters} of the experiment  in Ref.~\cite{experiment}, we show the resulting
entropy as a function of the temperature in Fig.~\ref{entropy}. We can see that it depends only weakly on the interaction $g_{BF}$ -- in stark contrast to the situation including the lattice, which we treat in the subsequent Section~\ref{lattice}. Furthermore, over the whole range of interactions, it is higher than the entropy for a purely bosonic system (note that this is not the same as the non-interacting situation, simply due to the fermionic contribution $s_F$ to the entropy of the mixture), for which the entropy may be obtained from the approximative  expression for the energy \cite{stringari_journal}
\begin{equation}
\begin{split}
\tfrac{E}{N_Bk_BT_c^0}=
at^4+b\bigl[N_B^{1/6}\tfrac{a_{BB}}{a_{ho}}(1-t^3)\bigr]^{2/5}(5+16t^3),
\end{split}
\end{equation}
which holds over a wide temperature and parameter range \cite{stringari_journal,stringari_review,adiabatic_bosons_proper}. Here,
$a=3\zeta(4)/\zeta(3)$, $b=\zeta(3)^{1/3}15^{2/5}/14$, $a_{ho}$ is the harmonic oscillator length of the trapping potential, and $T_c^0$ the critical temperature for bose condensation in the absence of interactions.  
This nourishes the hope of being able to also arrive at a closed expression for the entropy of the mixture, which is however beyond the scope of this work.

\section{Deep optical lattice}
\label{lattice}
For sufficiently deep optical lattices, the system may be described \cite{gap} by the single-band Bose-Fermi-Hubbard Hamiltonian \cite{albus},
$\hat{H}=\hat{J}+\hat{H}_0$,
\begin{equation}
\begin{split}
\hat{J}&=-J_B\sum_{\langle i,j\rangle}\hat{b}_i^\dagger\hat{b}_j-J_F\sum_{\langle i,j\rangle}\hat{f}_i^\dagger\hat{f}_j,\\
\hat{H}_0&=\sum_{i}\left[U\hat{n}_i(\hat{n}_i-1)/2+V\hat{n}_i\hat{m}_i-\mu_i\hat{n}_i-\nu_i\hat{m}_i\right],
\end{split}
\end{equation}
where $\hat{n}_i=\hat{b}_i^\dagger\hat{b}_i$, $\hat{m}_i=\hat{f}_i^\dagger\hat{f}_i$, the $\hat{b}_i$ ($\hat{f}_i$)
are bosonic (fermionic) annihilation operators, and the site-dependent chemical potentials, $\mu_i=\mu-t_i^B$, $\nu_i=\nu-t_i^F$,
account for the trapping potentials $t_i^{B/F}$ and control the number of particles. 
This model is obtained by including the lattice into the microscopic model in Eq.~(\ref{microscopic}), expanding the field operators in the Wannier basis, and neglecting all bands above the lowest band \cite{gap} and contributions beyond nearest neighbours.
The amplitudes $J_{B/F}, U, t_i^{B/F}, V\propto g_{BF}\propto a_{BF}$ are then obtained from a single-particle band-structure calculation for appropriate lattice parameters \cite{parameters} and the overlap of  the resulting Wannier functions. For a discussion of the validity of the single band approximation and contact interaction, we refer the reader to Refs.~\cite{mering,buechler_ww,saenz_trap,saenz_single_site,self_trapping}.

Similar to Ref.\ \cite{adiabatic1}, we obtain the entropy as a function of temperature employing the 
thermodynamic interaction picture, treating $\hat{J}$ as a perturbation up to first order (the difference to Ref.\ \cite{adiabatic1} being that we do not employ
a local density approximation and do not take geometrical averages of the trapping frequencies), which is applicable for $1\gg J_{B/F}\beta$ ($\simeq~\text{nK}/T$ in the situation at hand) \cite{landau}. This amounts to approximating
$\me^{-\beta\hat{H}}\approx \me^{-\beta\hat{H}_0}(\id-\hat{\gamma})$, $\hat{\gamma}=\int_0^\beta\md x\,\me^{x\hat{H}_0}\hat{J}\me^{-x\hat{H}_0}$. For given temperature this yields the partition function $Z=\text{tr}[\me^{-\beta\hat{H}}]\approx \text{tr}[\me^{-\beta\hat{H}_0}(\id-\hat{\gamma})]$ and the total number of bosons and fermions as a function of 
the chemical potentials. Numerically solving $\langle\sum_i\hat{n}_i\rangle=N_B$, $\langle\sum_i\hat{m}_i\rangle=N_F$
for $\mu,\nu$ then yields the entropy $S(T)=\beta\langle\hat{H}\rangle+\log(Z)$ for
given temperature and particle numbers $N_{B/F}$ up to first order in $\hat{J}$. Note that we consider the full three-dimensional anisotropic experimental situation, i.e., we need to take a large number of lattice sites into account, which results in quite involved numerics.

Fig.~\ref{entropy} summarizes the result of this procedure for the experimental parameters of Ref.~\cite{experiment}. The most prominent feature of $S(T)$ is its strong dependence on the Bose-Fermi interaction $V\propto a_{BF}$, while we found only a weak dependence in Section~\ref{no_lattice}. For fixed temperature, starting from a plateau for strong attraction, the entropy
increases until it reaches a maximum around $a_{BF}=0$, from which it decreases with increasing $a_{BF}$  to 
a plateau for strong repulsion. It is this behaviour that will crucially influence the temperature $T_f$ in the lattice and hence also the coherence properties of the bosonic atoms, which displays the same strong dependence on the Bose-Fermi interaction, see Section~\ref{section_visi}.
 The plateaus for large $|a_{BF}|$ are easily explained: For large repulsion, phase separation occurs and once this phase is entered, increasing $a_{BF}$ further does not have any effect. For large attraction on the other hand, bosons and fermions are forced to occupy the same lattice sites,
and again further increasing $|a_{BF}|$ does not have any effect.
Comparing the entropy of the mixture to the purely bosonic situation (note that this is not the same as the non-interacting case as, of course, also for $a_{BF}=0$, the fermions contribute to the total entropy), we see that the former is always higher than the latter for the considered parameter regime. As we will see in the subsequent Section, while
the adiabatic ramping up of the lattice leads to adiabatic cooling,
this causes the mixture to be less cooled than bosons would be without fermions.

\section{The temperature in the lattice}
\begin{figure}
\includegraphics[width=0.85\columnwidth]{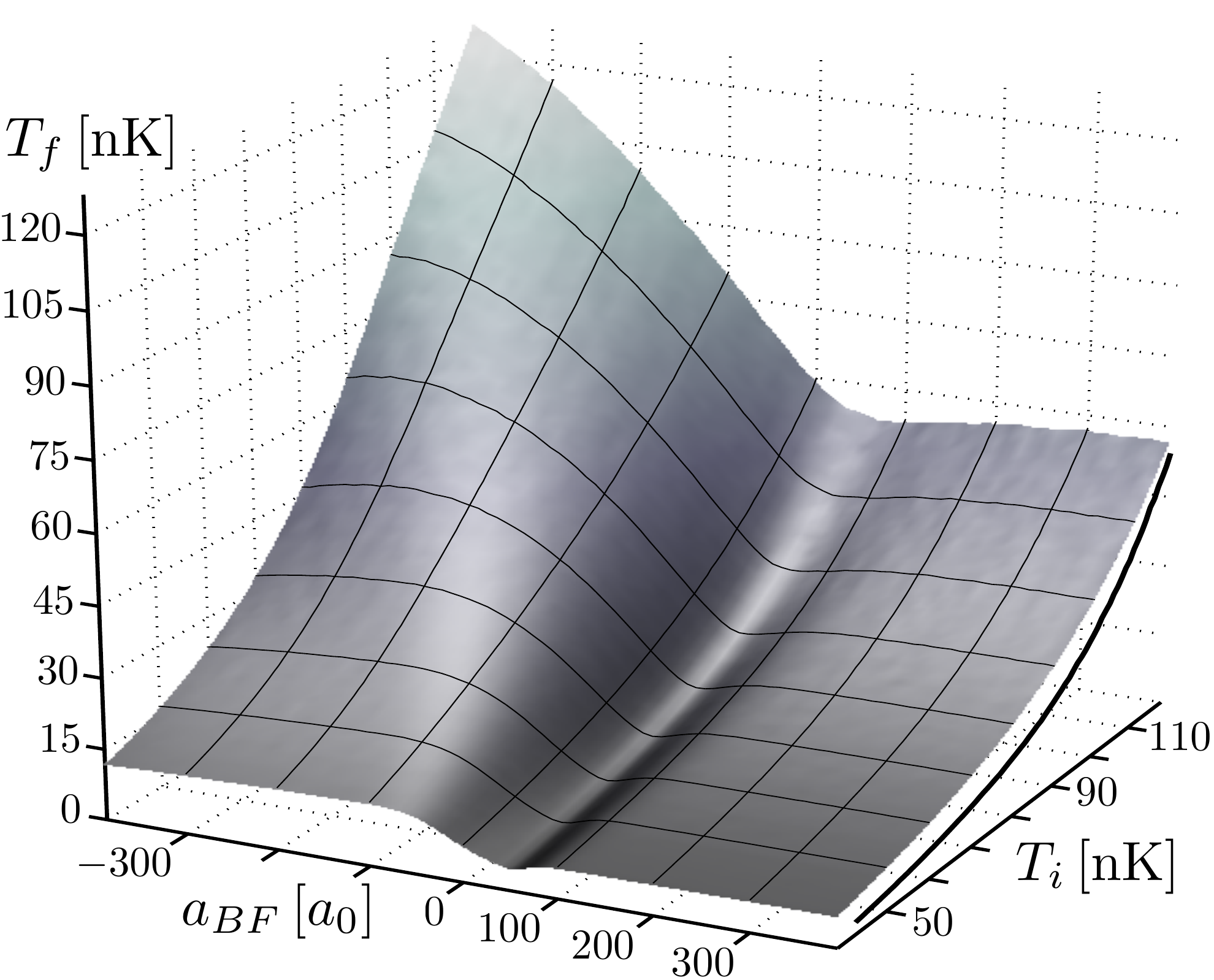}
\caption{
\label{Tfin}
Temperature of the mixture in an optical lattice (obtained by entropy matching from the data in Fig.~\ref{entropy}) as a function of the Bose-Fermi scattering lenght $a_{BF}$ ($\propto g_{BF}\propto V$, the Bose-Fermi interaction) and the initial temperature without lattice $T_i$. The bold line at $a_{BF}=400$ shows the same for a purely bosonic system, for which the final temperature is of course independent of $a_{BF}$.}
\end{figure}
Having obtained the entropies with, $S_{f}$, and without, $S_{i}$, lattice, we are now in the position
to obtain the temperature in the lattice, $T_f$, for given initial temperature, $T_i$, by matching
the respective entropies $S_f(T_f)=S_i(T_i)$. If the optical lattice is indeed raised adiabatically and the mixture is in thermal equilibrium, this enables us to compute $T_f$ as a function of $T_i$, which can be measured as, without lattice, thermometry methods are well established. Fig.~\ref{Tfin} shows the result obtained by matching the entropies in Fig.~\ref{entropy}. As $S_f$ in Fig.~\ref{entropy} already suggests, we find a strong dependence
of the temperature in the lattice on the Bose-Fermi interaction, most pronounced for high initial temperatures.
The qualitative behaviour of $T_f$ is similar to $S_f$: For fixed $T_i$ starting from a plateau at large attraction, the temperature decreases with increasing $a_{BF}$, reaches a minimum around $a_{BF}=0$,
from which it increases with increasing $a_{BF}$ to a plateau at high repulsion. We also depict $T_f$
for a purely bosonic system, which shows that while over most of the parameter regime, the mixture is cooled, the cooling is less than it would be without fermions. Being equipped with $T_f$, we can now move on to study the dependence of the bosonic coherence on the Bose-Fermi interaction.

\section{Visibility of the bosonic quasi-momentum distribution}
\label{section_visi}
\begin{figure}
\includegraphics[width=0.85\columnwidth]{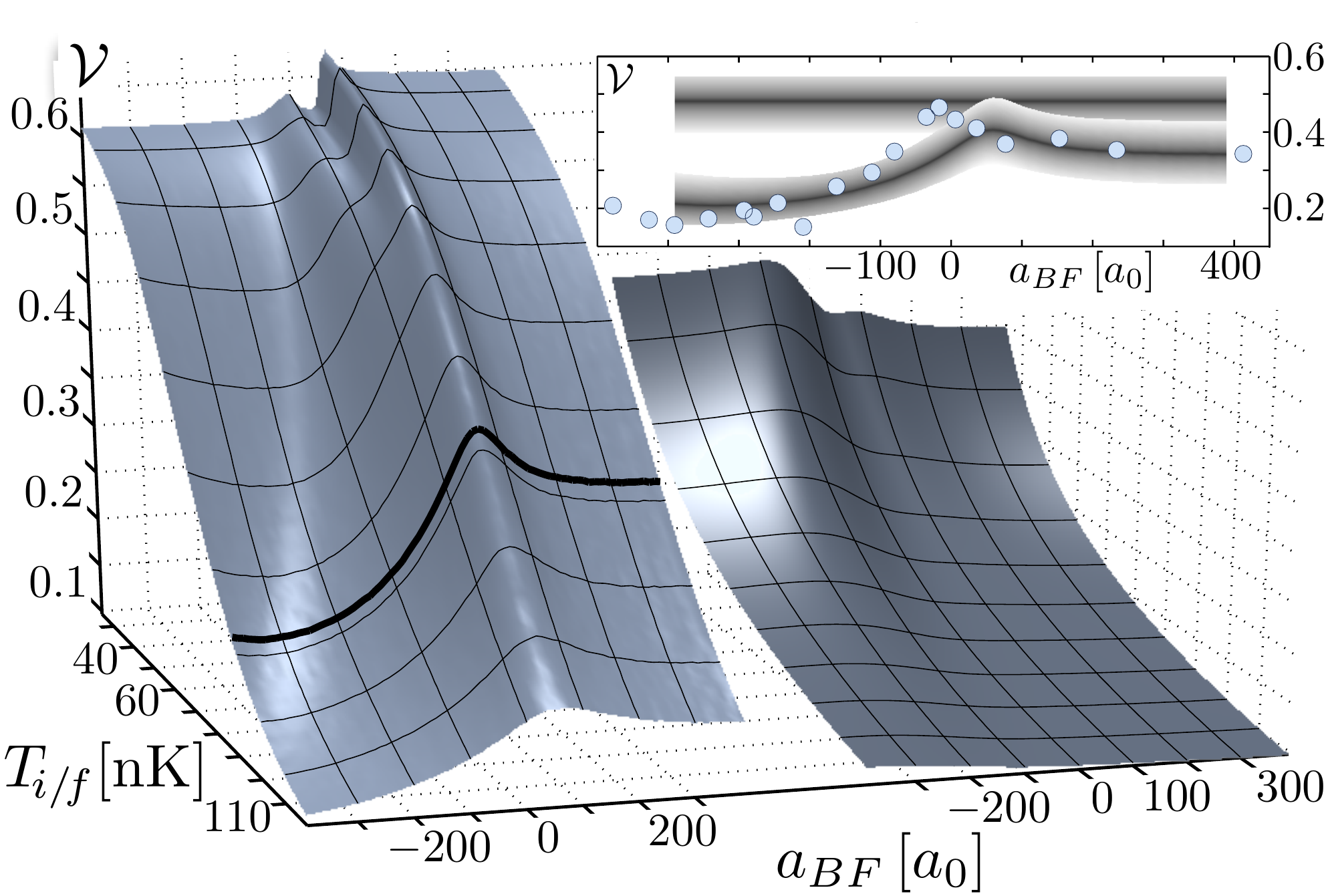}
\caption{\label{visi}
Visibility, $\mathcal{V}$, of the time-of-flight distribution at a lattice depth of twelve recoil energies as a function of the Bose-Fermi scattering length $a_{BF}$ ($\propto g_{BF}\propto V$, the Bose-Fermi interaction) and initial temperature without the lattice $T_i$ (left surface, $T_f$ obtained as in Fig.~\ref{Tfin}) and temperature in the lattice $T_f$ (right surface). Inset shows the same 
at the initial temperature indicated by the bold line
on the left surface for a purely bosonic system (including an uncertainty of 10\% in the initial temperature)
and the mixture (including an uncertainty of 15\% in the initial temperature). The initial temperature was chosen such that the purely bosonic situation matches the experiment \cite{experiment}. Circles are the data points from said experiment. The mixture consists of $4\times 10^5$ $^{87}$Rb and $3\times 10^5$ $^{40}$K atoms and all parameters (trap geometry, lattice depth) are as in this experiment and there are no free parameters in the theory.
}
\end{figure}
After switching off all potentials and letting the atom cloud evolve freely for a time $t$, the density of bosons is well approximated by \cite{stringari_qmd,bloch_qmd}
\begin{equation}
\begin{split}
n(\vec{p})\propto
\sum_{\vec{i},\vec{j}}\langle\hat{b}_{\vec{i}}^\dagger\hat{b}_{\vec{j}}\rangle
\me^{\mi\vec{p}a(\vec{i}-\vec{j})}\me^{\mi\frac{1}{4\tau} (|\vec{j}|^2-|\vec{i}|^2)}\hspace{1cm}\\
\times w(\vec{p}-\tfrac{\vec{i}}{2a\tau})w(\vec{p}-\tfrac{\vec{j}}{2a\tau}),
\end{split}
\end{equation}
where $\tau=\hbar t/(2m_Ba^2)$ and $w$ is the Fourier transform of the Wannier function centered at zero. This density is measured by taking an absorption image of the cloud, resulting in the column density 
 $n(p_x,p_y)=\int\!\md p_z\,n(\vec{p})$. For shallow lattices and low temperatures, i.e., in the superfluid regime, this density displays a pronounced interference pattern, which vanishes deep in the Mott regime (ultra-deep lattices) and for high temperatures. Hence, the visibility of this interference pattern, $\mathcal{V}=(n_{max}-n_{min})/(n_{max}+n_{min})$ \cite{visi_note}, is an indicator for the coherence of the $^{87}$Rb atoms. We calculate $\mathcal{V}$ 
 for a given temperature and number of particles by computing the two-point correlations $\langle\hat{b}_{\vec{i}}^\dagger\hat{b}_{\vec{j}}\rangle$ in thermal perturbation theory up to first order in $\hat{J}$, $\langle\hat{b}_{\vec{i}}^\dagger\hat{b}_{\vec{j}}\rangle\approx
 \text{tr}[\me^{-\beta\hat{H}_0}(\id-\hat{\gamma})\hat{b}_{\vec{i}}^\dagger\hat{b}_{\vec{j}}]/ \text{tr}[\me^{-\beta\hat{H}_0}(\id-\hat{\gamma})]$. From this we obtain the column density $n(p_x,p_y)$ and the visibility $\mathcal{V}$ \cite{visi_note} up to first order in $\hat{J}$. The result of this computation is shown in Fig.~\ref{visi} for parameters as in Ref.~\cite{experiment,parameters} and for two different temperatures: We depict $\mathcal{V}$ as a function of the temperature in the lattice $T_f$ (right surface) and as a function of the initial temperature $T_i$ (left surface). This corresponds to two different scenarios: (a) A scenario in which adiabatic cooling mechanism are omitted and hence 
the temperature in the lattice $T_f$ does not depend on the Bose-Fermi interaction and (b) the more realistic scenario in which the final temperature does depend on $a_{BF}$ through the entropy matching described above. As Fig.~\ref{visi} shows, the two scenarios result in opposed behaviours of $\mathcal{V}$ as a function of the interaction strength $a_{BF}$. For all temperatures and large $|a_{BF}|$, the visibility $\mathcal{V}$ is higher on the attractive side than on the repulsive side of the interaction for scenario (a), while we see the exact opposite for scenario (b).

\subsection{Comparison to the experiment}
 
Before we compare our results to the experiment, we recall the assumptions and approximations of our theory. We have computed the temperature in the lattice by matching the entropies of the mixture
with and without lattice (under the isentropic assumption, i.e., under the assumption that the lattice is ramped up adiabatically, which was recently confirmed for $^{87}$Rb atoms \cite{quantum_simulator}, this indeed yields the temperature in the lattice as a function of the initial temperature). To obtain the latter, we invoked the Hartree-Fock-Bogoliubov-Popov mean-field approximation, and computed the former within thermal perturbation theory up to first order in the tunnelling parameter of the single-band Bose-Fermi-Hubbard model (note that while the influence of higher bands in the lattice can not be completely ruled out and can have an effect on the bosonic coherence \cite{self_trapping,mering},
we are in a regime in which their occupation is expected to be small \cite{gap}). For all calculations, we assumed the mixture to be in thermal equilibrium such that we can assign one temperature to the mixture. 
This assumption is well justified if the mixture has sufficient time to equilibrate and gets worse  the smaller the Bose-Fermi interaction. Finally, we computed the visibility of the time-of-flight interference pattern, within
thermal perturbation theory up to first order in the tunnelling parameter.

The inset of Fig.~\ref{visi} compares our results directly to the experiment in Ref.~\cite{experiment}. We assume that the initial temperature for all measurements was approximately the same as in the purely bosonic situation. We can see that the results display the same qualitative behaviour: Starting at strong attraction, the visibility decreases to a minimum, from which it increases with increasing $a_{BF}$ up to a maximum at around $a_{BF}=0$, and finally decreases to a plateaux for strong repulsion. This is in stark contrast to what one finds without taking intrinsic temperature effects into account (see right surface in Fig.~\ref{visi}): The visibility
would, at the relevant temperatures, simply decrease monotonically with increasing interaction.

Fig.~\ref{visi} also shows a shift of the theoretical results relative to the experimental data.
All our results are parameterized by the Bose-Fermi scattering length $a_{BF}$, which is tuned in the experiment by addressing the magnetic Feshbach resonance at around $B_0\approx 546.9\,$G \cite{feshbach_experiment_1,feshbach_experiment_2}.
Close to resonance, magnetic field $B$ and scattering length are related by
\begin{equation}
\label{eq_feshbach}
a_{BF}=a_{BF}^{(0)}(1-\tfrac{\Delta B}{B-B_0}),
\end{equation}
where the resonance is at $B_0$, $a_{BF}^{(0)}$ is the background scattering length, and $\Delta B$ the width of the resonance. A faithful experimental determination of all the parameters in this relation is an extremely difficult endeavour. They depend on external parameters such as the trapping potential \cite{saenz_trap} -- and, even more so, the tight ``trapping" within a lattice site \cite{buechler_ww,saenz_single_site}. Especially the latter complicates a direct comparison to the experiment: Neither ab initio calculations for realistic inter-atomic potentials nor experimental measurements of the dependence of $a_{BF}$ on the magnetic field are available for the situation at hand. In fact, the experimental visibility shown in Fig.~\ref{visi} is really a function of the magnetic field, with $a_{BF}$ determined from Eq.~(\ref{eq_feshbach}) with $B_0= 546.9\,$G, $\Delta B=-2.9\,$G, and $a_{BF}^{(0)}=-185a_0$ \cite{thorsten}. Of course, due to the mentioned difficulties, this scattering length is a priori not the same as the one used to model the inter-atomic  contact potential. Exploring wether this may explain the remaining discrepancy between theory and experiment in Fig.~\ref{visi} is an exciting theoretical and experimental challenge that might result in a better understanding of the dependence of $a_{BF}$ on external potentials. We hope that, due to the  pronounced features of the
visibility---most prominently the location of the maximum---the present study can contribute to the work along this direction.

\section{Conclusion}
We have studied intrinsic temperature effects in Bose-Fermi mixtures taking the full three-dimensional anisotropic experimental situation into account. Under the adiabatic assumption, we have determined the temperature in the lattice as a function of the temperature
before the lattice is ramped up and found a strong dependence on the inter-species interaction.
This dependence affects the coherence of the bosons and is displayed in the visibility of the time-of-flight interference pattern, which we have compared to the experiment in Ref.~\cite{experiment}, finding qualitative agreement. Not including these temperature effects results in a very different dependence on the interaction and lead us to conclude that they need to be incorporated into any realistic description of Bose-Fermi mixtures in optical lattices. The remaining discrepancy between theory and experiment could be attributed to the dependence of the inter-species scattering length on external (trapping) potentials and we hope that it motivates further investigations of said dependence.

\section{Acknowledgements}
We greatfully acknowledge fruitful discussions with J.~Eisert, P.~Ernst, S.~G\"otze, M.~Ohliger, C.~Ospelkaus, S.~Ospelkaus, A.~Saenz,
P.~Schneider,
K.~Sengstock 
 and thank Th.~Best for helpful discussions concerning the experiment and for providing the experimental data.


\begin{thebibliography}{99}
\bibitem{parameters} Without lattice, the trap frequencies are given by $\omega^B_{\text{radial}}=33\,\text{Hz}$, $\omega^B_{\text{axial}}=113\,\text{Hz}$,  $\omega^F_{\text{radial}}=45\,\text{Hz}$, $\omega^F_{\text{axial}}=194\,\text{Hz}$. The lattice introduces an additional confinement, leading to $\omega^B_{\text{radial}}=29\,\text{Hz}$, $\omega^B_{\text{axial}}=112\,\text{Hz}$,  $\omega^F_{\text{radial}}=34.5\,\text{Hz}$, $\omega^F_{\text{axial}}=192\,\text{Hz}$
at a lattice depth of twelve recoil energies \cite{thorsten}. We consider the experiment with $N_B=4\times 10^5$, and $N_F=3\times 10^5$. The optical lattice in Ref.~\cite{experiment} was operated at $755\,$nm and chosen such that the lattice depths, in units of their respective recoil energies, were  the same for both species.
\bibitem{gap} At the considered lattice depth of $12E^{B/F}_R$, $E_R^{B/F}=(2\pi\hbar/\lambda)^2/(2m_{B/F})$, the gap between the first two lowest bands is
 $\Delta E_{B/F}\approx 6.1E^{B/F}_R$ and the interaction energies and temperatures $V/\Delta E_F\approx 0.0004a_{BF}/a_0$, $U/\Delta E_B\approx 0.057$, $k_B/\Delta E_B\approx 0.0008/\text{nK}$. Hence, all interaction energies and temperatures considered here are much smaller than the band gap. Furthermore, both Refs.~\cite{bfh_exp1,bfh_exp2} report that neither bosons nor fermions occupied higher bands.
\bibitem{visi_note} To minimize noise, one averages over a box in reciprocal space around $n_{max}=n(2\pi/a,0)$, $n_{min}=n(\sqrt{2}\pi/a,\sqrt{2}\pi/a)$, which we take into account.
\end{thebibliography}
\end{document}